\documentclass[jsl]{asl}
\usepackage{mathrsfs}

\title[Unary primitive recursive functions]
{Unary primitive recursive functions}


\author{Daniel E. Severin}
\revauthor{Severin, Daniel E.}

\address{Facultad de Ciencias Exactas, Ingenieria y Agrimensura\\
Universidad Nacional de Rosario\\
Rosario, Santa Fe, Argentina}

\urladdr{http://www.fceia.unr.edu.ar/$\sim$daniel}

\email{daniel@fceia.unr.edu.ar}

\newtheorem{Theorem}{Theorem}[section]
\newtheorem{Lemma}[Theorem]{Lemma}

\theoremstyle{definition}
\newtheorem{Definition}[Theorem]{Definition}
\newtheorem{Remark}[Theorem]{Remark}

\renewcommand{\qedsymbol}{$\square$}
\def \sub { \textsf{subst} }
\def \clos { \textsf{clos} }
\def \Prim { \textsf{Prim} }
\def \UPrim { \textsf{Prim}(\nat, \nat) }
\def \nat { \mathbb{N} }
\def \to { \rightarrow }
\def \module { {\rm mod} }
\def \monus { \dotminus }
\def \ufrp { \mathscr{F} }
\def \closf { \clos\ufrp }
\def \la { \langle }
\def \ra { \rangle }
\def \sq { \square }
\def \Hf { H\!f }
\def \Mod { M\!od }
\def \df { \equiv }
\def \alg { \mathscr{A} }
\def \rec { \mathscr{R} }
\def \mix { \mathscr{M} }
\def \xx { \mathfrak{x} }
\def \Z { \overline{0} }
\def \X { \overline{1} }
\def \Y { \overline{2} }
\def \MRO { \!\!$\textsf{rec}_1$\! }
\def \PRO { \!\!$\textsf{rec}_2$\! }

\def \PIO { \!\!$\textsf{rec}_4$\! }
\def \MIZ { \!\!$\textsf{rec}_5$\! }
\def \PIZ { \!\!$\textsf{rec}_6$\! }
\def \MI { \!\!$\textsf{rec}_7$\! }
\def \PI { \!\!$\textsf{rec}_8$\! }

\begin{document}

\begin{abstract}
In this article, we study some new characterizations of primitive recursive
functions based on restricted forms of primitive recursion, improving the
pioneering work of R. M. Robinson and M. D. Gladstone. We reduce
certain recursion schemes (mixed/pure iteration without parameters) and we
characterize one-argument primitive recursive functions as the closure under
substitution and iteration of certain optimal sets.

\end{abstract}

\maketitle


\section{Introduction.} $\Prim$, i.e. the set of \emph{primitive recursive
functions}, is the closure under substitution and primitive recursion of
zero, successor and projection functions. For a detailed definition, the
reader is referred to any standard work, for instance chapter 1 of \cite{GOO57}.
A suitable subset is $\UPrim$, i.e. the set of \emph {unary primitive recursive
functions}. It will be one of the objects of our research.

Recursion schemes have been studied intensively during the twentieth century.
In particular, R. M. Robinson\cite{ROB47, ROB55B} and his wife
J. Robinson\cite{ROB50, ROB55A} proved that it is sufficient to consider
one-argument functions because functions of several arguments can be reduced to
them using pairing strategies. Later on, Gladstone\cite{GLA67, GLA71}
and Georgieva\cite{GEO76A} made improvements to the recursion schemes.
At the same time as the study of recursive functions, several classifications were
carried out over $\UPrim$. The first one was Grzegorczyk hierarchy\cite{GRZ53}.
Since then, other hierarchies have appeared (cf. \cite{RIT65,AXT66,GEO76B,DIM81,NAU83}).
Finally, some algebraic properties of $\UPrim$ were verified in \cite{SZA85}.
Similar topics are covered in \cite{GM88,MAZ05}.

The present paper improves the work of Robinson\cite{ROB47} and Gladstone\cite{GLA71}.

The paper is organized as follows: In \S\ref{SNOT} we will give a useful
symbolic notation for writing functions. In \S\ref{SPRE} we will show previous
results, and the facts to be proved here. In \S\ref{SMIX} we will analyze a
possible reduction in one of the recursion schemes. More precisely, mixed
iteration without parameters with $a$ fixed is as expressive as mixed iteration
without parameters with $a$ variable (the meaning of these schemes and $a$ can
be found in \S\ref{SPRE}). In \S\ref{SITD} we will do the same thing with pure
iteration without parameters. And, in \S\ref{SIUO} we will characterize unary
primitive recursive functions as the closure of the set including
$x \longmapsto 1$ and $x \longmapsto x - \lfloor \sqrt{x} \rfloor^2$ with
respect to substitution, iteration and the following operator:
$f \longmapsto f + I$, where $I$ is the identity function on natural numbers.


\section{Notation.} \label{SNOT} To denote arbitrary functions we
shall use letters in uppercase such as $F$, $G$ and $H$. To
denote natural variables we shall use $x$, $y$, $z \ldots$, whereas
$a$, $b$, $\ldots$ are used to denote constants. Throughout
the paper, the following functions will be used:
\begin{itemize}
	\item Basic functions:
\begin{align*}
	I(x) &= x ~ &\text{(identity)} \\
	\overline{n}(x) &= n ~ &\text{(constants)} \\
	S(x) &= x + 1 ~ &\text{(successor)} \\
	P(x) &= x \monus 1 ~ &\text{(predecessor)} \\
	I^n_k(x_1, x_2, \ldots, x_n) &= x_k, \textrm{~for~} 1\!\leq\!k\!\leq\!n
		 ~ &\text{(projections)}
\end{align*}
	\item Arithmetic functions:
\begin{align*}
	D(x) &= 2x ~ &\text{(double)} \\
	Sq(x) &= x^2 ~ &\text{(square)} \\
	\Hf(x) &= \lfloor x / 2 \rfloor ~ &\text{(half)} \\
	Pw(x) &= 2^x ~ &\text{(power of two)} \\
	Rt(x) &= \lfloor \sqrt{x} \rfloor
		~ &\text{(integer square root)}
\end{align*}
	\item Cantor pairing functions:
\begin{align*}
	A(x) &= \lfloor (x^2 + x) / 2 \rfloor
		~ &\text{($x$-th triangular number)} \\
	V(x) &= \bigg\lfloor \dfrac{\lfloor \sqrt{8x + 1}
		\rfloor - 1}{2} \bigg\rfloor
		~ &\text{(inverse of $A$)} \\
	J(x, y) &= A(x + y) + x ~ &\text{(pairing function)} \\
	K(x) &= x - A(V(x)) ~ &\text{(first inverse)} \\
	L(x) &= A(V(x) + 1) - x - 1 ~ &\text{(second inverse)}
\end{align*}
	\item Binary functions:
\begin{align*}
	x \monus y &= \begin{cases}
		x - y & \text{if $x \geq y$} \\
		0 & \text{otherwise}
	\end{cases} ~ &\text{(arithmetic difference)} \\
	|x - y| &= \begin{cases}
		x - y & \text{if $x \geq y$} \\
		y - x & \text{otherwise}
	\end{cases} ~ &\text{(distance)}
\end{align*}
	\item Other functions:\footnote{Some authors write $0^x$,
$\overline{\textup{sg}}(x)$ or $\textup{cosg}(x)$ instead of
$O(x)$.}
\begin{align*}
	O(x) &= \begin{cases}
		1 & \text{if $x = 0$} \\
		0 & \text{otherwise}
		\end{cases} ~ &\text{(power of zero, cosignum)} \\
	Sgn(x) &= \begin{cases}
		0 & \text{if $x = 0$} \\
		1 & \text{otherwise}
		\end{cases} ~ &\text{(signum)} \\
	N(x) &= x ~\module~ 2
		~ &\text{(characteristic of odd numbers)} \\
	E(x) &= x - \lfloor \sqrt{x} \rfloor^2
		~ &\text{(excess over a square)} \\
	Q(x) &= \begin{cases}
		1 & \text{if $x$ is a square} \\
		0 & \text{otherwise}
		\end{cases}
		~ &\text{(characteristic of square numbers)}
\end{align*}
\end{itemize}
Let $F$, $G$, $G_1$, $\ldots$, $G_m$ be functions, and
$\xx = (x_1, x_2, \ldots, x_n)$, i.e. a $n$-tuple. The following operators
on natural number functions will be used:
\begin{itemize}
	\item Substitution:
\[ \sub(F, G_1, G_2, \ldots, G_m)(\xx) =
	F(G_1(\xx), G_2(\xx), \ldots, G_m(\xx)). \]
	A more special case is defined for one-argument functions,
\[ (F \circ G)(\xx) = F(G(\xx)), \]
\[ (F G)(x) = F(G(x)). \]
	\item Primitive recursion:
\begin{align*}
	\rec[F, G](\xx, 0) &= F(\xx),\\
	\rec[F, G](\xx, y + 1) &= G(\xx, y, \rec[F, G](\xx, y)).
\end{align*}
	\item Restricted forms of primitive recursion:\footnote{Notations
$F^\sq$ and $F^{\sq(a)}$ are due to Szalkai\cite{SZA85}.}
\begin{align*}
	1)~ \mix[F](0) &= 0,\\
	\mix[F](x + 1) &= F(x, \mix[F](x)),\\
	2)~ F^{\sq(a)}(0) &= a,\\
	F^{\sq(a)}(x + 1) &= F(F^{\sq(a)}(x)).\\
	3)~ F^\sq(x) &= F^{\sq(0)}(x).
\end{align*}
	\item Power:
\begin{align*}
	F^0(x) &= x,\\
	F^{n + 1}(x) &= F(F^n(x)).
\end{align*}
	\item Miscellaneous:
\begin{align*}
	1)~ (F + G)(x) &= F(x) + G(x),\\
	2)~ (F \monus G)(x) &= F(x) \monus G(x),\\
	3)~ |F - G|(x) &= |F(x) - G(x)|, \\
	4)~ J(F, G)(x) &= J(F(x), G(x)).
\end{align*}
\end{itemize}

In order to decrease the size of this article and improve readability,
we will give a symbolic notation for representing functions. If the
definition of a new function $F : \nat^n \to \nat$ is
\[ F(x_1, x_2, \ldots, x_n) =
	\textsf{expression}(x_1, x_2, \ldots, x_n), \]
we will write
\[ F ~\df~ \textsf{expression}, \]
where $\textsf{expression}$ is composed by the functions and
operators previously defined. Precedence and associativity rules are
shown in table 1.
\begin{table}[tb]
\caption{Precedence and associativity of operators.}
\begin{center}
\begin{tabular}{|l|l|l|}
\hline
Precedence & Operators & Associativity \\
\hline
\hline
First & $F + G$, $F \monus G$ & Left \\
\hline
Second & $F G$, $F \circ G$ & Any \\
\hline
Third & $F^n$, $F^\sq$, $F^{\sq(a)}$ & $\relbar$ \\
\hline
\end{tabular}
\end{center}
\end{table}
Here are some examples of well-formed expressions:
\begin{align*}
	&D \df \mix[S \circ S \circ I^2_2],
	&O \df \sub(\rec[\X, P \circ I^3_3], I, I), \\
	&Pw \df S(I + I + \X)^\sq,	&V \df \Hf~P~Rt~S~D~D~D.
\end{align*}

A finite set of initial functions and of functional operators is called \emph{basis}.
We will denote with
\[ \ufrp = \la F_1, F_2, \ldots, F_n, F^\oplus, \ldots,
	F \otimes G, \ldots \ra \]
the basis composed by the initial functions $F_1$, $F_2$, $\ldots$,
$F_n$, the unary operators $F^\oplus$, $\ldots$, the binary operators
$F \otimes G$, $\ldots$ and so on.

We will denote with $\closf$ the closure of the basis $\ufrp$. An example is
\[ \Prim = \clos\la \Z, S, I^n_k, \sub, \rec[F, G] \ra. \]


\section{Preliminaries.} \label{SPRE} In \cite{ROB47}, some recursion schemes are
introduced (all of them are particular cases of primitive recursion):
\begin{enumerate}
	\item Mixed recursion with one parameter:\\
		$F(x, 0) = G(x),~~~ F(x, y + 1) = H(x, y, F(x, y))$.
	\item Pure recursion with one parameter:\\
		$F(x, 0) = G(x),~~~ F(x, y + 1) = H(x, F(x, y))$.
	\item Mixed iteration with one parameter:\\
		$F(x, 0) = x,~~~ F(x, y + 1) = H(y, F(x, y))$.
	\item Pure iteration with one parameter:\\
		$F(x, 0) = x,~~~ F(x, y + 1) = H(F(x, y))$.
	\item Mixed iteration without parameters:\\
		$F(0) = a,~~~ F(y + 1) = H(y,F(y))$.
	\item Pure iteration without parameters:\\
		$F(0) = a,~~~ F(y + 1) = H(F(y))$.
	\item Mixed iteration without parameters,
		and $a = 0$:\\
		$F(0) = 0,~~~ F(y + 1) = H(y,F(y))$.
	\item Pure iteration without parameters,
		and $a = 0$ (or simply called \emph{iteration}):\\
		$F(0) = 0,~~~ F(y + 1) = H(F(y))$.
\end{enumerate}
We will refer to these schemes as \MRO, \PRO, \ldots, \PI in the same
order listed above.
Note that schemes \MRO, \MI and \PI have symbolic notations:
$F \df H^{\sq(a)}$, $F \df \mix[H]$ and $F \df H^\sq$.

Robinson and Gladstone proved that the primitive recursion scheme can be
replaced by one of the cases with one parameter, i.e.
\MRO-\PIO. They also proved that the cases without parameters,
i.e. \MIZ-\PI, are adequate but certain functions must be added to
the initial functions. Table 2 summarizes which functions are sufficient to be
included as initial functions (the symbol $\relbar$ denotes the null set).
\begin{table}[tb]
\caption{Table of functions that must be added as initial functions.}
\begin{center}
\begin{tabular}{|c|c|c|c|c|}
\hline
 & \multicolumn{2}{c|}{} & \multicolumn{2}{c|}{} \\
 & \multicolumn{2}{c|}{One Parameter}
 	& \multicolumn{2}{c|}{No Parameter} \\
 & \multicolumn{2}{c|}{} & \multicolumn{2}{c|}{} \\
\cline{2-5}
 & & & & \\
 & Recursion & Iteration & $a$ variable & $a=0$ (fixed) \\
 & & & & \\
\hline
 & & & & $x+y$, $Q$ \cite{ROB47} \\
Mixed & $\relbar$ \cite{ROB47} & $\relbar$ \cite{ROB47}
	& $x+y$ \cite{GLA71} & $|x-y|$ \cite{ROB47} \\
 & & & & $x+y$, $O$ \S\ref{SMIX} \\
\hline
 & & & & $x+y$, $E$ \cite{ROB47} \\
 & & & & $x+y$, $K$ \cite{ROB55B} \\
Pure & $\relbar$ \cite{GLA67} & $\relbar$ \cite{GLA71}
	& $|x-y|$ \cite{GLA71} & $x+y$, $L$ \cite{ROB55B} \\
 & & & & $J$, $K$ \cite{ROB55B} \\
 & & & $x \monus y$ \cite{GEO76A,NAU83}
	& $J$, $L$ \cite{ROB55B} \\
 & & & & $|x-y|$ \S\ref{SITD} \\
 & & & & $x \monus y$ \S\ref{SITD} \\
\hline
\end{tabular}
\end{center}
\end{table}
In this table, the references indicate where the proofs of previous
results can be found and the section references indicates where are
the proofs of new results. Now, the tables that appeared on
p. 929 of \cite{ROB47} and on p. 654 of \cite{GLA71} can be substituted
by our table.

In the cases without parameters, it is not necessary to take zero
function as an initial function because it can be obtained from identity
and iteration as follows:
\[ \Z(0) = 0,~~~ \Z(x + 1) = I(\Z(x)). \]
Moreover, in the pure cases without parameters, it is not
necessary to take projection functions as initial functions if we are
considering one-argument functions. In $\UPrim$, there is only one projection,
the identity function, which can be obtained from successor and iteration
as follows:
\[ I(0) = 0,~~~ I(x + 1) = S(I(x)). \]
Notice that all constant functions belong to every case given in table 2,
since they can be generated using zero and successor functions:
$\overline{n}(x) = S^n(\Z(x))$. Constant functions of more that one
argument can be defined composing a one-argument constant function with
an arbitrary function of $n$ arguments (e.g. a projection).\\

At the end of \S4 of \cite{ROB47}, Robinson determined that $Sq$, $O$,
$\Hf$, $Rt$, addition and substraction\footnote{The notation $x-y$ without dot
or vertical bars, will always be used in an ambiguous sense, to stand
for any function $F(x, y)$ which is equal to $x-y$ for $x\!\geq\!y$, regardless
of its value when $x\!<\!y$. Any difference function, such as
$x \monus y$ or $|x-y|$, can substitute $x-y$.} ($x-y$) are sufficient to
add as initial functions when we works with \MIZ-\PI. We will rewrite this
result in the next lemma.

\begin{Lemma} \label{ROBINSON1} For $i \in \{ 5, 6, 7, 8\}$,
\[ \Prim = \clos \la S, I^n_k, Sq, O, \Hf, Rt, +, -, \sub, \textsf{rec}_i \ra. \]
\end{Lemma}

In some sections, we just will work with unary primitive recursive
functions and the scheme of iteration. The following definition
will help us.

\begin{Definition}
We say that a basis $\ufrp$ is \emph{suitable} when $\clos \ufrp = \UPrim$.
\end{Definition}

\begin{Lemma} \label{ROBINSON2}
The basis $\la S, Sq, O, \Hf, Rt, F+G, F-G, FG, F^\sq \ra$ is suitable.
\end{Lemma}
\begin{proof}
It follows from Lemma \ref{ROBINSON1} (also see Theorem 2 of \cite{ROB47})
and $I \df S^\sq$. Due to the impossibility of introducing binary functions,
we must incorporate operators such as $F+G$ and $F-G$.
\end{proof}

Furthermore, Robinson proved that $\Prim$ can be obtained by adding
projection functions, addition and the substitution operator to $\UPrim$
(cf. \S7 of \cite{ROB47}). We write this result as another lemma.

\begin{Lemma} \label{ROBINSON3} Let $\ufrp$ be a suitable basis. Then,
$\Prim = \clos (\ufrp + \la I^n_k, +, \sub \ra)$.
\end{Lemma}

Now, we will derive a list of suitable bases for the pure cases without
parameters (see table 3, the format is the same as in table 2).
Bases provided in \S$\ref{SIUO}$ are simpler than Robinson's bases.
In fact, the successor can be substituted by $\X$, and the addition operator
can be substituted by a unary operator of the form $f \longmapsto f + I$.
\begin{table}[tb]
\caption{Initial functions for characterizations of $\UPrim$ using pure iteration.}
\begin{center}
\begin{tabular}{|c|c|}
\hline
 & \\
$a$ variable & $a=0$ (fixed) \\
 & \\
\hline
 & $S$, $E$, $F+G$ \cite{ROB47} \\
 & $S$, $K$, $F+G$ \cite{ROB55B} \\
 & $S$, $L$, $F+G$ \cite{ROB55B} \\
$S$, $|F-G|$ \cite{GLA71}		& $S$, $E$, $J(F,G)$ \cite{ROB55B} \\
 & $S$, $K$, $J(F,G)$ \cite{ROB55B} \\
 & $S$, $L$, $J(F,G)$ \cite{ROB55B} \\
 & $S$, $|F-G|$ \S\ref{SITD} \\
$S$, $F \monus G$ \cite{GEO76A,NAU83}	& $S$, $F \monus G$ \S\ref{SITD} \\
 & $\X$, $E$, $F+I$ \S\ref{SIUO} \\
 & $\X$, $K$, $F+I$ \S\ref{SIUO} \\
 & $\X$, $L$, $F+I$ \S\ref{SIUO} \\  
\hline
\end{tabular}
\end{center}
\end{table}


\section{Mixed iteration without parameters.} \label{SMIX}
In \S4 of \cite{GLA71}, Gladstone showed that \MIZ is
adequate if we include the addition function. Our aim is to
verify that \MI is adequate too, but we
must incorporate a function that is not non-decreasing: cosignum.
In order to do this, we need to follow the same steps as
\cite{GLA71} but keeping in mind that we must use \MI.

At the scope of this section, let $\ufrp = \la S, I^n_k, O, +, \sub, \mix[F] \ra$.

\begin{Lemma} \label{LEMMAPRED} $P, N, D, Sq, \Hf, Pw \in \closf$.
\end{Lemma}
\begin{proof}
In the first place, $P \df \mix[I^2_1]$, $N \df \mix[O \circ I^2_2]$ and
$D \df \mix[S \circ S \circ I^2_2]$. Furthermore, we have:
\begin{itemize}
	\item Square: $Sq(0) = 0, Sq(x + 1) = Sq(x) + 2x + 1.$\\
	$Sq \df \mix[\sub(+, S \circ I^2_2, D \circ I^2_1)].$
	\item Half: $\Hf(0) = 0, \Hf(x + 1) = \Hf(x) + N(x).$\\
	$\Hf \df \mix[\sub(+, I^2_2, N \circ I^2_1)].$
	\item Power of two: Let $F$ be defined as follows: $F(0)=0$,
	$F(x+1)=2F(x)+1$. Therefore, $F(x)=2^x-1$ and
	$Pw \df S \circ \mix[S \circ D \circ I^2_2].$
\end{itemize}
\end{proof}

\begin{Lemma} \label{LEMMADELTA}
The function $\delta(x, y) = \begin{cases}
	1 & \text{if $x = y$} \\
	0 & \text{otherwise}
\end{cases}$ (namely \emph{Kronecker delta function}) belongs to $\closf$.
\end{Lemma}
\begin{proof}
In Lemma 6 of \S4 of \cite{GLA71}, the following function $f$ is
defined using scheme \MIZ:
\begin{align*}
	f(0) &= 2, \\
	f(x + 1) &= N(z) + z + 2^{x + O(N(z))} + 2^{x + 2O(N(z))},
\end{align*}
where $z = \bigg\lfloor \dfrac{f(x)}{2} \bigg\rfloor$.
We can \emph{simulate} this function by transferring the index
in one unit:
\begin{align*}
	f'(0) &= 0, \\
	f'(x + 1) &= N(z') + z' + 2^{x + O(N(z')) - 1}
		+ 2^{x + 2O(N(z')) - 1} + O(O(x)),
\end{align*}
where $z' = \bigg\lfloor \dfrac{f'(x) - 1}{2} \bigg\rfloor$.
Thus, $f(x) = f'(x + 1) - 1$.\\
Now, let $g$ be defined as $g(0)=0$,
$g(x + 1) = N\bigg\lfloor \dfrac{f(x - 1)}{2} \bigg\rfloor$.\\
According to Gladstone,\footnote{In his paper, $g(x)$ returns 0
when $x$ is a power of two, and 1 if not.}
\[ g(x) = \begin{cases}
	1 & \text{if $x$ is a power of two}, \\
	0 & \text{otherwise}.
\end{cases}\]
Note that $x = y$ iff $2^x + 2^y$ is a power of two,
so $\delta(x, y) = g(2^x + 2^y)$.
\end{proof}

\begin{Lemma} \label{LEMMASQROOT} $Rt, - \in \closf$.
\end{Lemma}
\begin{proof}
Integer square root is computed as follows: $Rt(0)=0$,
$Rt(x+1)=Rt(x)+\delta((Rt(x)+1)^2,x+1)$. Symbolically,
\[ Rt \df \mix[\sub(+, I^2_2, \sub(\delta,
	Sq \circ S \circ I^2_2, S \circ I^2_1))]. \]
Let $H$ be defined as follows:
\begin{align*}
	H(0) &= 0, \\
	H(x + 1) &= H(x) + 2N(\lfloor \sqrt{x} \rfloor) \monus 1.
\end{align*}
Hence, $H(x) = E(x)$ when $\lfloor \sqrt{x} \rfloor$ is an odd
number (cf. part (4) of \S6 of \cite{ROB47}), so that
\[ x - y = H((2x + 2y)^2 + 5x + 3y + 1) \]
whenever $x \geq y$. The formula above defines the function substraction as
a functional operator. Finally, $- \df I_1^2 - I_2^2$.
\end{proof}

\begin{Theorem} \label{THEOREMMIX}
$\Prim = \clos \la S, I^n_k, O, +, \sub, \mix[F] \ra$.
\end{Theorem}
\begin{proof}
It follows from Lemma \ref{ROBINSON1} and
Lemmata \ref{LEMMAPRED}-\ref{LEMMASQROOT}.
\end{proof}

We will prove two theorems which explain the reason
we included cosignum function in Theorem \ref{THEOREMMIX}.

\begin{Theorem}
Let $\ufrp' = \ufrp - \la O \ra$ (the result of removing $O$ from the basis $\ufrp$).
Every function $F$ of one argument of $\closf'$ is non-decreasing:
\[ \forall_{x \in \nat} ~~~ F(x) \leq F(x+1). \]
\end{Theorem}
\begin{proof}
We will proceed by structural induction over functions defined
using one argument. The fact is trivial for identity and
successor function. If $F$ and $G$ are non-decreasing functions,
its substitution (i.e. $F \circ G$) and its addition
(i.e. $\sub(+, F, G)$) are non-decreasing too.\\
Now, let $F$ be defined as
\[ F(0) = 0,~~~ F(x + 1) = G(x,F(x)). \]
Clearly, $G$ is a function written in terms of $I^2_1$, $I^2_2$ and
non-decreasing functions. So, $G$ satisfy the following property:
\[ \forall_{a,b,x,y \in \nat} ~~~ G(x, y) \leq G(x + a, y + b). \]
Suppose that $F(x) \leq F(x + 1)$. Then,
\[ G(x, F(x)) \leq G(x + 1, F(x + 1)). \]
Therefore, $F(x + 1) \leq F(x + 2)$.
\end{proof}

\begin{Theorem} \label{THEOREMSGN}
$\Prim = \clos \la S, I^n_k, \hat{F}, +, \sub, \mix[F] \ra$,
where $\hat{F}$ is not non-decreasing.
\end{Theorem}
\begin{proof}
If $\hat{F}$ is not non-decreasing then exists a natural number $a$ that
verifies $\hat{F}(a) > \hat{F}(a + 1)$. Let $G$ be defined as $G(x) = \hat{F}(x + a)$,
i.e $G \df \hat{F} \circ S^a$. Thus, $G(0) > G(1)$. Let $H$ be defined as
$H(x) = G(x) \monus G(1)$, i.e. $H \df P^{G(1)} \circ G$, where
$P \df \mix[I^2_1]$. Thus, $H(0) > H(1) = 0$.\\
Next, let $Sgn \df \mix[~\X~]$. It follows easily that
\begin{align*}
	Sgn(H(Sgn(0))) &= Sgn(H(0)) = 1, \\
	Sgn(H(Sgn(x + 1))) &= Sgn(H(1)) = 0.
\end{align*}
Therefore, $O \df Sgn \circ H \circ Sgn$. And now, we can apply Theorem \ref{THEOREMMIX}.
\end{proof}

\begin{Remark} \label{REMARK}
In this section, we fixed the value of $a$ to zero. However, we
could have fixed the value of $a$ to another number.\footnote{This
differs from Gladstone, because he used \MIZ with several values
of $a$ (more precisely, with $a \in \{ 0, 1, 2 \} $). We show that it is
sufficient to choose one value for $a$.}\\
We will show that scheme \MI can be expressed using
\MIZ with $a > 0$. First, we define the functions below:
\begin{align*}
	&\hat{P}(0) = a,~~~ \hat{P}(x + 1) = x,
		~~~\textrm{i.e.}~~~ \hat{P} \df \mix_a[I^2_1],\\
	&\overline{a}(0) = a,~~~ \overline{a}(x + 1) = \overline{a}(x),
		~~~\textrm{i.e.}~~~ \overline{a} \df \mix_a[I^2_2],\\
	&\Z \df \hat{P}^a \circ \overline{a},\\
	&\hat{O}(0) = a,~~~ \hat{O}(x + 1) = 0,
		~~~\textrm{i.e.}~~~ \hat{O} \df \mix_a[\Z],
\end{align*}
where $\mix_a[F](0) = a$, $\mix_a[F](x + 1) = F(x, \mix_a[F](x))$.\\
Now, every function $F$ which satisfies $F(0) = 0$ and
$F(x + 1) = H(x, F(x))$ will be written as follows:
\begin{align*}
	G(0) &= a, \\
	G(x + 1) &= H(x, G(x) - a) + a.
\end{align*}
By a simple induction, $F(x) = G(x) - a$, and
\[ \mix[H] \df \hat{P}^a \circ \mix_a[S^a \circ \sub(H, I^2_1, \hat{P}^a \circ I^2_2)]. \]
Note also that $\hat{O}$ is not non-decreasing.
So, applying Theorem \ref{THEOREMSGN} we prove
that $\Prim = \clos\la S, I^n_k, +, \sub, \mix_a[F] \ra$.
\end{Remark}


\section{Iteration and difference.} \label{SITD} We will follow \S5
of \cite{GLA71} (also see Lemma 1 of \cite{NAU83}), replacing \PIZ
by \PI: $\Prim$ is generated using a difference function (may
be $|x-y|$ or $x \monus y$) as the unique initial function. However,
we will propose an equivalent statement. Let $\ufrp$ be
$\la S, |F - G|, FG, F^\sq \ra$ or $\la S, F \monus G, FG, F^\sq \ra$.
Our intention is to prove that $\ufrp$ is suitable.

As much as possible, we will try to use $F-G$ instead of $|F-G|$ and
$F \monus G$, but taking care of not subtracting two functions that render
the expression meaningless. In the first place,
\begin{align*}
	&I \df S^\sq,		&D \df (SS)^\sq, \\
	&\Z \df S - S,		&\X \df S \Z, \\
	&Pw \df S(SD)^\sq,	&Sgn \df \X^\sq, \\
	&P \df I - Sgn,		&O \df \X - Sgn.
\end{align*}

Next step is to construct the addition. The following sequence of functions
\[ \begin{cases}
	f_0 &\df S, \\
	f_{n+1} &\df f_n^{\sq(f_n(1))}
\end{cases} \]
is a kind of Ackermann's sequences (i.e. if $f(x,n) = f_n(x)$ then $f$
grows faster than any primitive recursive function; nevertheless, $f_n$
is primitive recursive). Georgieva\cite{GEO76A} discovered a method
for constructing the addition between two functions, based on this sequence.\\
Let $F, G \in \closf$. According to Lemma 6 of \cite{GEO76A},
there exists $i \in \nat$ such that $F(x) \leq f_i(x)$ for
every $x$ (and there exists $j \in \nat$ such that
$G(x) \leq f_j(x)$). Let $k$ be the maximum value between $i$
and $j$. Hence, $F(x) + G(x) \leq 2f_k(x)$. Therefore (cf.
Lemma 7 of \cite{GEO76A}),
\[ F + G \equiv D f_k - ((D f_k - F) - G). \]
Now, we will explain how to construct $f_i$ given $F$ by means
of the following recursive definition:
\begin{align*}
\begin{cases}
	\alg : \ufrp \to \nat & \\
	\alg(S) &= 0 \\
	\alg(|F - G|) &= \max(\alg(F), \alg(G)) \\
	\alg(F \monus G) &= \alg(F) \\
	\alg(FG) &= \max(\alg(F), \alg(G)) + 2 \\
	\alg(F^\sq) &= \alg(F) + 1
\end{cases}
\end{align*}

To express $F + G$ using \PI instead of \PIZ, we need
only to generate a sequence that grows faster than $f_n$.
\begin{Lemma} \label{LEMMAB}
The following sequence of functions
\[ \begin{cases}
	B_0 &\df S, \\
	B_{n+1} &\df (S^{f_n(1)}~B_n)^\sq
\end{cases} \]
satisfies
\[ \forall_{x,n \in \nat} ~~~ B_n(x + 1) \geq f_n(x). \]
\end{Lemma}
\begin{proof}
First, we will try to rewrite $f_n$ with iterations.
\[ \begin{cases}
	f'_0(x) &= x, \\
	f'_{n+1}(0) &= 0, \\
	f'_{n+1}(x + 1) &= g_n(f'_{n+1}(x))
\end{cases} \]
where $g_n(x) = f_n(1) O(x) + f'_n(x) Sgn(x)$. Hence, $f'_n(x + 1) = f_n(x)$
(by a simple induction on $x$ and $n$).
Consider the sequence
\[ \begin{cases}
	B_0(x) &= x + 1, \\
	B_{n+1}(0) &= 0, \\
	B_{n+1}(x + 1) &= h_n(B_{n+1}(x))
\end{cases} \]
where $h_n(x) = f_n(1) + B_n(x)$. Clearly, $B_{n+1}(x) \geq f'_{n+1}(x)$
if $B_n(x) \geq f'_n(x)$ (by comparing $g_n$ and $h_n$). We conclude that
$B_n(x + 1) \geq f_n(x)$.
\end{proof}

\begin{Lemma}
If $F, G \in \closf$ then $F + G \in \closf$.
\end{Lemma}
\begin{proof}
Remember that there exists $i, j \in \nat$ such that
$F(x) \leq f_i(x)$ and $G(x) \leq f_j(x)$. By virtue of the
previous lemma, $F(x) \leq B_i(x + 1)$ and $G(x) \leq B_j(x + 1)$.
Let $k = \max(i, j)$, so
\[ F(x) + G(x) = 2 B_k(x + 1) - ((2 B_k(x + 1) - F(x)) - G(x)). \]
In other words,
\[ F + G \equiv D B_{\max(\alg(F),\alg(G))} S -
	((D B_{\max(\alg(F),\alg(G))} S - F) - G). \]
\end{proof}

Now, we only need to prove that $Sq, Rt, \Hf \in \closf$. We will do this
in the next lemmata.

\begin{Lemma}
The following families of functions belong to $\closf$:
\begin{itemize}
	\item Characteristic of $n$:
		\[ O_n(x) = \begin{cases}
			1 & \text{if $x = n$}, \\
			0 & \text{otherwise}.
		\end{cases} \]
	\item Multiplication functions:
		\[ M_n(x) = nx. \]
	\item Cycle functions:
		\[ C_{n+2}(x) = \begin{cases}
		x + 1 & \text{if $x \leq n$}, \\
		0 & \text{otherwise}.
		\end{cases} \]
	\item Moduli functions:
		\[ \Mod_{n+2}(x) = x ~\module~ (n + 2). \]
	\item Division functions:
		\[ Div_{n+2}(x) = \lfloor x / (n + 2) \rfloor. \]
\end{itemize}
\end{Lemma}
\begin{proof}
We will show the formulas of each one in the same order. All of them
can be proved easily by induction on $n$.\footnote{Some proofs can be
consulted in \S5 of \cite{GLA71}.}\\
Characteristic of $n$:
\[ O_0 \df O,~~~ O_1 \df O(O + P),~~~ O_{n+2} \df O_{n+1} P. \]
Multiplication functions:
\[ M_n \df (S^n)^\sq. \]
Cycle functions:
\[ C_2 \df O,~~~ C_{n+3} \df C_{n+2} + M_{n+2} O_{n+1}. \]
Moduli functions:
\[ \Mod_{n+2} \df C_{n+2}^\sq. \]
Division functions:
\[ Div_{n+2} \df (S + O~\Mod_{n+3}~S~S)^\sq - I. \]
\end{proof}

\begin{Definition} \label{GUARDA}
The conditional operator $F \to G$ is defined as follows
\[ (F \to G)(x) = \begin{cases}
	G(x) & \text{if $F(x) = 0$}, \\
	0 & \text{otherwise}.
\end{cases} \]
\end{Definition}

\begin{Lemma}
If $F, G \in \closf$ then $F \to G \in \closf$.
\end{Lemma}
\begin{proof}
Let $\alpha(x) = 2^{x+1+\Mod_2(x)} - 2^{x+1}$. If $x$ is even,
$\alpha(x) = 0$. And if $x$ is odd, $\alpha(x) = 2^{x+1}$.
In formal terms,
\[ \alpha \df Pw(S + \Mod_2) - Pw~S. \]
Now, we will divide the proof in two cases depending on the substraction
operator which we are working:
\begin{itemize}
	 \item Distance: Let $\beta \df (|\alpha - (I + Pw)| + I) - Pw$.
	If $x$ is even,	$\beta(x) = 2x$. And if $x$ is odd, $\beta(x) = 0$.
	\item Arithmetic difference: Let $\beta \df D \monus \alpha$.
	If $x$ is even, $\beta(x) = 2x$. And if $x$ is odd, $\beta(x) = 0$.
\end{itemize}
Finally, we will observe the behavior of $w = \beta(2z + Sgn(y))$.
When $y$ is zero, $w = 4z$. And when $y$ is positive, $w = 0$.
So, $w = 4.(F \to G)(x)$ if $y = F(x)$ and $z = G(x)$, and
\[ (F \to G) \df Div_4 \beta (DG + Sgn~F). \]
\end{proof}

\begin{Lemma}
$Q \in \closf$.
\end{Lemma}
\begin{proof}
We follow Lemma 2.3 of \cite{GEO76B}. Let $W$ be defined as follows:
\[ W(x) = \begin{cases}
	2 & \text{if $x = 0$}, \\
	\lfloor 3x / 2 \rfloor & \text{if $x \neq 0$,
		$x ~\module~ 10 = 0$}, \\
	\lfloor 2x / 5 \rfloor & \text{if $x \neq 0$,
		$x ~\module~ 2 \neq 0$, $x ~\module~ 5 = 0$}, \\
	\lfloor 2x / 3 \rfloor & \text{if $x \neq 0$,
		$x ~\module~ 3 = 0$, $x ~\module~ 5 \neq 0$}, \\
	\lfloor 15x / 2 \rfloor & \text{if $x \neq 0$,
		$x ~\module~ 3 \neq 0$, $x ~\module~ 5 \neq 0$}. \\
\end{cases} \]
For all $x > 0$, $W^\sq(x) ~\module~ 3 \neq 0$ if and only if $x$
is a square.
To write $W$ we must use the operator defined above (see \ref{GUARDA}):
\begin{align*}
	W_1(x) &\df DO, \\
	W_2(x) &\df (O + \Mod_{10} \to Div_2~M_3), \\
	W_3(x) &\df (O + O~\Mod_2 + \Mod_5 \to Div_5~D), \\
	W_4(x) &\df (O + \Mod_3 + O~\Mod_5 \to Div_3~D), \\
	W_5(x) &\df(O + O~\Mod_3 + O~\Mod_5 \to Div_2~M_{15}).
\end{align*}
Each $W_i$ represents one case (one line of the definition of $W$).
The conditions are mutually exclusive, so $W(x) = W_i(x)$ for some
$i$ between 1 and 5.\\
Thus, $W \df W_1 + W_2 + W_3 + W_4 + W_5$ and
\[ Q \df Sgn~\Mod_3~W^\sq + O. \]
\end{proof}

\begin{Lemma} \label{LEMMAFIN}
$Sq, Rt, \Hf \in \closf$.
\end{Lemma}
\begin{proof}
We follow \S5 of \cite{ROB47}. Suppose that
$R(x) = x + 2 \lfloor \sqrt{x} \rfloor$.\\
Then, $R \df (S + DQSSSS)^\sq$ and
\[ Sq \df (SR)^\sq, Rt \df Div_2(R-I), \Hf \df Div_2. \]
\end{proof}

\begin{Theorem} \label{THEORDIFF}
The bases $\la S, |F - G|, FG, F^\sq \ra$ and $\la S, F \monus G, FG, F^\sq \ra$
are both suitable.
\end{Theorem}
\begin{proof}
It follows from Lemma \ref{ROBINSON2} and Lemmata \ref{LEMMAB}-\ref{LEMMAFIN}.
\end{proof}

We conclude this section with the following theorem, which is a consequence
of the previous results.

\begin{Theorem} \label{THEORDIFF2}
$\Prim = \clos \la S, I^n_k, \ominus, \sub, F^\sq \ra$, where $\ominus(x, y)$ can be
$x \monus y$ or $|x - y|$.
\end{Theorem}
\begin{proof}
Let $\ufrp = \la S, F \ominus G, FG, F^\sq \ra$. In virtue of Theorem \ref{THEORDIFF}, $\ufrp$ is
suitable, and therefore the operator addition belongs to $\closf$. Note that
$+ \df I^2_1 + I^2_2$, $F \ominus G \df \sub(\ominus, F, G)$ and $FG \df \sub(F, G)$, so
$\clos \la S, I^n_k, \ominus, \sub, F^\sq \ra \supseteq
\clos \la S, I^n_k, F \ominus G, \sub, F^\sq \ra \supseteq
\clos (\ufrp + \la I^n_k, \sub \ra) \supseteq
\clos (\ufrp + \la I^n_k, +, \sub \ra) = \Prim$ (use Lemma \ref{ROBINSON3}).
Therefore, $\Prim = \clos \la S, I^n_k, \ominus, \sub, F^\sq \ra$.
\end{proof}


\begin{Remark} \label{REMPIZ}
In this section, we used scheme \PIZ with $a = 0$.
However, we could have fixed the value of $a$ to another number as we
did in \ref{REMARK}. In fact,
\[ F^\sq \df \hat{P}^a (S^a F \hat{P}^a)^{\sq(a)}\]
where $\hat{P}$ may be $|S - \Y|$ or $S \monus \Y$, and $\Y \df SS(S-S)$.
\end{Remark}

\begin{Remark}
A further line of inquiry is to analyze if it is possible to rewrite the
operator $F \to G$ using the difference $F - G$ instead of $F \monus G$
and $|F - G|$. For example, if we prove that $Sq \in \closf$, then we can write
\[ (F \to G) \df Div_2(Sq(OF + G) - (Sq~O~F) - (Sq~G)) \]
and replace $F \monus G$ and $|F - G|$ by $F - G$ in table 3.
\end{Remark}


\section{Iteration and unary operator.} \label{SIUO}
Robinson\cite{ROB47} proved that $\la S, E, F+G, FG, F^\sq \ra$ is a suitable
basis. In this section, we will simplify this result, showing that 
$\la \X, E, F^+, FG, F^\sq \ra$ is suitable too.
Let $\ufrp = \la \X, E, F^+, FG, F^\sq \ra$, where the operator $F^+$ is
defined as
\[ F^+(x) = F(x) + x, \]
and its precedence is the same as in $F^\sq$.
The aim of this section is to show that it is not necessary to have a
binary operator such as the addition (except for, of course, the substitution).
In fact, the addition can be replaced by a unary operator.

In the first place, the following functions belong to $\closf$:
\begin{align*}
	&S \df \X^+,			&Sgn \df \X^\sq, \\
	&\Z \df E \X, 			&Mod_3 \df (ESS)^\sq, \\
	&D \df (\Z^+)^+,		&M_3 \df D^+, \\
	&O \df E~D~S~Sgn,		&Q \df OE, \\
	&R \df ((DQSSS)^+ S)^\sq,	&Sq \df (SR)^\sq.
\end{align*}

\begin{Definition}
In this section, the following operators on one argument functions will be used:
\begin{align*}
	&F^-(x) = F(x) - x,			&(F \otimes G)(x) = F(x)G(x), \\
	&(F \oplus G)(x) = F(x) + G^2(x),	&(F \ominus G)(x) = F(x) - G^2(x),
\end{align*}
whenever $F(x) \geq x$ and $G(x) \geq x$ for every $x \in \nat$.
In the definition of $\ominus$, $F(x) \geq G^2(x)$ must hold
too.\footnote{$G^2(x)$ must be read as $G(x)G(x)$,
and not $G(G(x))$.}\\
The precedences of $\oplus$ and $\ominus$ are the same as in the
addition, while the precedence of $F^-$ is the same as in $F^+$.
The precedence of $\otimes$ is between addition and substitution
(like products in arithmetical expressions).

\end{Definition}

\begin{Lemma} \label{LEMMASUBSTR}
Let $F \in \closf$. If $F(x) \geq x$ for every $x$, then
$F^- \in \closf$.
\end{Lemma}
\begin{proof}
Robinson has proved that, if $\alpha \geq \beta$, then
\[ \alpha - \beta = E((\alpha + \beta)^2 + 3\alpha + \beta + 1). \]
If we take $\alpha = F(x)$ and $\beta = x$, the formula becomes
\[ F(x) - x = E((F(x) + x)^2 + 3 F(x) + x + 1). \]
The following diagram shows how to compute $F^-$:
\[ x \mapstochar\xrightarrow{(Sq~F^+)^+} F^+(x)^2 + x
	\mapstochar\xrightarrow{(M_3 F E)^+} F^+(x)^2 + 3 F(x) + x
	\mapstochar\xrightarrow{E S} F^-(x) \]
since $E((F(x) + x)^2 + x) = x$. Therefore,
\[ F^- \df E S (M_3 F E)^+ (Sq~F^+)^+. \]
\end{proof}

\begin{Lemma}
$\Hf, Rt \in \closf$.
\end{Lemma}
\begin{proof}
Cf. \S5 of \cite{ROB47}:
\begin{align*}
	\Hf &\df ((S~{Mod_3}^+)^\sq)^-, \\
	Rt &\df \Hf~R^-.
\end{align*}
\end{proof}

\begin{Lemma}
Let $F, G \in \closf$. If $F(x) \geq x$ and $G(x) \geq x$, then
$F \oplus G \in \closf$. If $F(x) \geq G^2(x)$ too, then
$F \ominus G \in \closf$.
\end{Lemma}
\begin{proof}
We will use the fact that if $G(x) \geq x$ then $E(G^2(x) + x) = x$.
Hence,
\begin{align*}
	&x \mapstochar\xrightarrow{(Sq~G)^+} G^2(x) + x
		\mapstochar\xrightarrow{(F^- E)^+} F(x) + G^2(x), \\
	&x \mapstochar\xrightarrow{(Sq~G)^+} G^2(x) + x
		\mapstochar\xrightarrow{(F^+ E)^-} F(x) - G^2(x).
\end{align*}
Therefore,
\begin{align*}
	F &\oplus G \df (F^- E)^+ (Sq~G)^+, \\
	F &\ominus G \df (F^+ E)^- (Sq~G)^+.
\end{align*}
\end{proof}

\begin{Lemma}
Let $F, G \in \closf$. If $F(x) \geq x$ and $G(x) \geq x$, then
$F \otimes G \in \closf$.
\end{Lemma}
\begin{proof}
Note that
\[ \alpha^2 \beta^2 = \dfrac{\bigl[ (\alpha^2 + 1)^2 + \beta^2
	- \alpha^4 - 1 \bigr]^2  - \beta^4}{4} - \alpha^4.\]
If we take $\alpha = F(x)$ and $\beta = G(x)$, we can reach
$\alpha \beta$ with
\[ F \otimes G \df Rt((\Hf~\Hf(Sq~\hat{P}((Sq~S~Sq~F \oplus G)
	\ominus Sq~F) ~\ominus~ Sq~G)) ~\ominus~ Sq~F), \]
where $\hat{P} \df ((Sq^+)^+)^+ \ominus S$, i.e. $\hat{P}(x + 1) = x$.
\end{proof}

\begin{Lemma} \label{LEMMALAST}
Let $F, G \in \closf$. So, $F + G \in \closf$. If $F(x) \geq G(x)$ for
every $x$, then $F - G \in \closf$ too.\footnote{The formula of $F - G$
works well even when $F(x) < G(x)$ for some values of $x$. We can use
them regardless of the values which render the formula meaningless.}
\end{Lemma}
\begin{proof}
First, we will compute the sum $F^+(x) + G^+(x)$ by using the following
properties: $(\alpha + \beta)^2 = 2\alpha\beta + \alpha^2 + \beta^2 $,
$F^+(x) \geq x$, $G^+(x) \geq x$,
\[ F^+(x) + G^+(x) =
	\sqrt{2 F^+(x) G^+(x) \oplus F^+(x) \oplus G^+(x) }. \]
Now, see that $F(x) + G(x) = F^+(x) + G^+(x) - 2x$. Therefore,
\[ F + G \df ((Rt((D(F^+ \otimes G^+) \oplus F^+) \oplus G^+))^-)^-. \]
To compute $F-G$, we can use the same trick as in Lemma
\ref{LEMMASUBSTR}. Finally,
\[ F - G \df ES(Sq(F + G) + M_3 F + G). \]
\end{proof}

\begin{Theorem} \label{THEOFINAL}
$\la \X, E, F^+, FG, F^\sq \ra$ is a suitable basis.
\end{Theorem}
\begin{proof}
It follows from Lemma \ref{ROBINSON2} and Lemmata \ref{LEMMASUBSTR}-\ref{LEMMALAST}.
\end{proof}

Now, we will prove that $\ufrp$ is suitable if we use $K$ (or $L$)
instead of $E$. Let $\ufrp' = \la \X, K, F^+, FG, F^\sq \ra$.
Following p. 664 of \cite{ROB55B},
\begin{align*}
	&S \df \X^+,			&Sgn \df \X^\sq, \\
	&\Z \df K \X,			&D \df (\Z^+)^+, \\
	&Y \df ((Sgn~K)^+ S)^\sq,	&Z \df (SSK^+)^\sq,
\end{align*}
where $Y(x) = 2x - \lfloor \sqrt{x} \rfloor$ and
$Z(x) = x(x + 3)/2$.\\
Let $F \in \closf'$. Using the fact that
\[ K((\alpha+\beta)(\alpha+\beta+3)/2+2\alpha+3)=\alpha-\beta,
\textrm{~if~} \alpha\!\geq\!\beta, \]
we see that
\[ I-F \df KSSS((ZF^+)^+)^+,
\textrm{~if~} x\!\geq\!F(x), \]
and that
\[ D-F \df KSSS(((((Z(F^+)^+)^+)^+)^+)^+),
\textrm{~if~} 2x\!\geq\!F(x). \]
Thus,
\[ Rt \df D - Y, Sq \df (((\X^+)^+ Rt)^+)^\sq, E \df I - Sq~Rt.\]
The application of Theorem \ref{THEOFINAL} makes $\ufrp'$ suitable.
If $\ufrp'' = \la \X, L, F^+, FG, F^\sq \ra$, we may define $K$ by
the formula $K \df L \X^+ ((\X^+)^+ L)^+$, and $\ufrp''$ results to be
suitable too.
\begin{Theorem} \label{THEOFINAL2}
$\la \X, K, F^+, FG, F^\sq \ra$ and $\la \X, L, F^+, FG, F^\sq \ra$ are suitable bases.
\end{Theorem}


\begin{Remark}
In this section, we used scheme \PIZ with $a = 0$. However, we
could have fixed the value of $a$ to another number as we did in
\ref{REMPIZ}. We only need a suitable predecessor $\hat{P}$,
i.e. $\hat{P}(x + 1) = x$.

For $a = 1$, we can proceed as follows:\\
(I) Let $\ufrp = \la E, F^+, FG, F^{\sq(1)} \ra$. Then,
\begin{align*}
	&O \df E^{\sq(1)},		&\Z \df EO, \\
	&\X \df O\Z,			&S \df \X^+, \\
	&D \df (\Z^+)^+,		&Q \df OE, \\
	&G \df (((DQSS)^+S)^{\sq(1)})^{\sq(1)},
	&\hat{P} \df ESS((G^+)^+)^+,
\end{align*}
where $G(x) = (x + 1)^2$. Now, iteration can be defined with
\[ F^\sq \df \hat{P} (S F \hat{P})^{\sq(1)}. \]
In virtue of Theorem \ref{THEOFINAL}, $\ufrp$ is suitable.\\
(II) Let $\ufrp = \la K, F^+, FG, F^{\sq(1)} \ra$. Then,
\begin{align*}
	&O \df K^{\sq(1)},		&\Z \df KO, \\
	&\X \df O\Z,			&S \df \X^+, \\
	&H \df S(S K^+ S)^{\sq(1)},	&\hat{P} \df KSSS(H^+)^+,
\end{align*}
where $H(x) = (x+1)(x+4)/2$. Now, iteration can be defined as in
(I). In virtue of Theorem \ref{THEOFINAL2}, $\ufrp$ is suitable.\\
(III) Let $\ufrp = \la L, F^+, FG, F^{\sq(1)} \ra$. Then,
\begin{align*}
	&\X \df L^{\sq(1)},	&K \df L \X^+ ((\X^+)^+ L)^+,
\end{align*}
and by using (II) we prove that $\ufrp$ is suitable.

For $a > 1$ it is not known if $\la X, F^+, FG, F^{\sq(a)} \ra$,
for $X \in \{E, K, L\}$, is suitable. This question will be
studied in the future.
\end{Remark}


\section{Acknowledgments.}
I would like to thank the referee for his helpful comments and suggestions.


\bibliographystyle{asl}
\bibliography{b}

\end{document}